# Effect of Strain on the Growth of InAs/GaSb Superlattices: An X-Ray Study

J.H. Li[1,2,+], D.W. Stokes[1,2], J.C. Wickett[1], O. Caha[1,*], K.E. Bassler[1,2], and S.C. Moss[1,2]

[1]Physics Department, University of Houston, Houston, TX 77204

[2] Texas Center for Superconductivity and Advanced Materials, University of Houston, Houston, TX 77204

## Abstract

We present a detailed x-ray diffraction study of the strain in InAs/GaSb superlattices grown by molecular beam epitaxy. The superlattices were grown with either InSb or GaAs interfaces. We show that the superlattice morphology, either planar or nanostructured, is dependent on the chemical bonds at the heterointerfaces. In both cases, the misfit strain has been determined for the superlattice layers and the interfaces. We also determined how the magnitude and sign of this strain is crucial in governing the morphology of the superlattice. Our analysis suggests that the growth of self-assembled nanostructures may be extended to many systems generally thought to have too small a lattice mismatch.

---

* On leave from the Institute of Condensed Matter Physics, Masaryk University, Brno, Czech Republic

* Current address: Rigaku Americas Corporation, The Woodlands, TX

I. Introduction

The formation of self-organized semiconductor epitaxial nanoscale structures (wells, wires and dots)[1,2] based on the morphological instability of strained molecular beam epitaxial (MBE) grown films, has been observed in many III-V systems and is of great importance from both a fundamental and technological point of view. Despite the progresses made towards understanding the formation of these nanostructures, there are still numerous obstacles which must be overcome before these structures may be utilized for practical applications. For example: (1) The epitaxial systems available for the self-assembled nanostructure formation is limited due to the requirement for the presence of large misfit strain between the layers. This requirement excludes many important optoelectronic materials, such as GaInP/GaAs and AlInAs/InP, which have misfits less than 1%, from being candidates for self-assembled nanostructure formation. (2) Once nanostructure formation has been established in a system, the ability to control and obtain uniform size and spatial distributions of the nanoscale assembly is challenging due to growth kinetics.

The self-assembled growth of epitaxial nanostructures, in principle, relies on the Stranski-Krastanov (SK) instability, where a film initially follows two-dimensional (2D) growth and then transitions into a three-dimensional (3D) growth due to elastic strain relaxation once the film thickness goes beyond a critical layer thickness. Thus, the SK instability is generally observed in heterostructures with a large misfit strain. The most studied systems, InAs/GaAs and Ge/Si, have misfits greater than 7% and 4%, respectively and result in the formation of quantum dots. However, with these large misfits, the self-assembled dots often exhibit nonuniform size and spatial distributions, which can only be improved by employing more complicated methods, such as surface patterning[3-8].

An unusual instability phenomena resulting in the self assembly of nanostructures has been observed in non-common anion III-V heterostructures with small misfit strains. Okada et al. observed that $In_{0.45}Ga_{0.55}As$/InP (001) with a small tensile strain of just +0.5% demonstrates morphological instability at a thickness of only a couple of monolayers (ML), where one monolayer is approximately 3 Å[9]. The formation of nanostructures at this critical thickness is comparable to the critical layer thickness at which nanostructures form during growth of InAs/GaAs heterostructures which have a compressive strain of about -7%. Similarly, Nosho et al. observed nanostructure formation in InAs/GaSb (001) superlattices with a small tensile strain of +0.62% also at a critical thickness of a few ML's[10].

In this article, we will present in-depth details associated with the experimental and theoretical analysis of planar (stable) and nanowire (unstable) layers in InAs/GaSb (001) superlattices. Nanostructure formation in this small-misfit system is atypical since this type of instability is generally observed in systems with large misfits. Key results of this study have also been presented elsewhere[11,12]. This study shows that the strain configuration of both the superlattice and interface layers determines the system's stability. This observation suggests the possibility of self-assembled nanostructure formation in other systems with small misfit, which may lead to the development of novel optoelectronic devices.

## II. Experiments

A series of $(InAs)_{13}/(GaSb)_{13}$ superlattices with 90-100 periods were grown by solid-source molecular beam epitaxy (MBE) on GaSb (001) or InAs (001) substrates using $As_4$ and $Sb_4$. Growth temperatures ranged from 295 to 335°C and the growth rate for both materials was 0.5 ML/s. Additional growth details are given elsewhere[13]. Since there are no common atoms across the heterointerfaces, the interfacial bonds between the layers can either be InSb or GaAs, as shown schematically in Fig. 1. However, by employing the growth technique, migration enhanced epitaxy (MEE), the composition of the interface (IF) layer can be controlled[10,13]. Table I gives information on four InAs/GaSb superlattices. Samples A and C were grown on GaSb (001) substrates with InSb interfacial bonds. Sample D was also grown on a GaSb (001) substrate, but with GaAs interfacial bonds and sample B was grown on an InAs (001) substrate with GaAs interfacial bonds.

Sample morphology was examined in ultrahigh vacuum by cross-sectional scanning tunneling microscopy (STM) at a constant current of 0.2 nA[13]. Samples grown with GaAs interfaces remained planar (stable), while those with InSb interfaces were nanowired (unstable). Reconstructed three-dimensional diagrams of the morphologies of nanowire sample C and planar sample D are shown in Fig. 2a and b, respectively. For the nanowire sample, the InAs layers (dark regions) undulate in thickness sinusoidally along the $[\bar{1}10]$ lateral direction, and are almost completely surrounded by the more uniform GaSb layers (bright regions). Moreover, the undulations of the neighboring InAs layers are out-of-phase, resulting in a nanowire array with a 2D centered rectangular (*cr*) symmetry. For the planar sample, a typical flat superlattice structure is observed. STM data was not taken for all samples; however, Fig. 2, is the typical structure seen for the planar and nanowired samples.

STM analysis does not reveal much information on the role of the interfaces in the instability; therefore, high resolution x-ray diffraction (XRD) was employed. From XRD measurements, strain and composition of the nanowire and planar superlattices were determined. Measurements were performed both in-house and at the National Synchrotron Light Source (NSLS) at the Brookhaven National Laboratory using four-circle diffractometers. In-house measurements were performed with a 12 kW Rigaku rotating anode x-ray source (Cu $K_{\alpha 1}$ radiation), while those at NSLS were performed with an x-ray energy of 8.0 keV (beamline X14A). In both cases, Si (111) monochromators were used. Sample and detector slits were chosen to attain a resolution necessary to separate the satellite peaks and maintain peak shapes. To determine if the sample had nanowire formation, as a first measure, line scans were performed in two sample azimuths, shown in Fig. 3, since the observed nanowire morphology is anisotropic. For azimuth 1, the plane of diffraction defined by the incident and exit x-ray wave vectors and the sample surface normal are parallel to the [110] direction and for azimuth 2, the plane of diffraction is set to be parallel to the $[\bar{1}10]$ direction.

## III. Experimental Results

The four samples listed in Table I were analyzed by XRD for both azimuths 1 and 2. For planar samples, B and D, both azimuths showed identical spectra indicating that there was no nanowire formation. Fig. 4 shows the measured (004) diffraction data (dots) and simulation (solid line) for samples B and D. From the interval between the satellite peaks, the real space superlattice periodicity was determined to be 82.4 Å and 80.2 Å, respectively. This is in excellent agreement with the expected period of 80 Å.

For the nanowired samples, reciprocal space maps (RSM) were taken for both azimuths and data is shown in Fig. 5 for sample A about the (002) reciprocal lattice point. It is clearly seen that when the plane of diffraction is parallel to [110], only 1D satellites in the Qz||[001] direction are observed. The intervals between the satellites (ΔQ=0.038 r.l.u.) give a vertical periodicity of 160 $\pm 3$ Å, which is twice as big as the designed 80 Å period. When the plane of diffraction is parallel to $[\bar{1}10]$, azimuth 2, 2D satellites in the ($Q_x$,$Q_z$) plane are observed with an *cr* symmetry, which is directly related to the symmetry of the nanowire arrays. The appearance of high-order satellites indicates that the nanowire array is highly ordered. From the intervals of the satellites of the nearest orders in the $Q_x$ and $Q_z$ direction, a lateral period of 1693 Å and a vertical period of 160 $\pm 3$ Å was determined. Similar diffraction patterns were also observed for nanowire sample C.

## IV. Theoretical Analysis

The lattice mismatch between InAs and GaSb is only 0.62%; therefore, the observed instability cannot be completely explained by the Stranski-Krastanov growth mode. In order to understand the instability observed for growth of the InAs/GaSb superlattices, a better insight of the strain and composition of the layers is necessary since strain is believed to be the driving force behind growth instability. Simulations of the experimental x-ray data, using a kinematic approach were used to extract the quantitative structural information for the samples.

A. Planar Samples

For planar samples, B and D, with GaAs interfacial bonds, the diffraction intensity is given as

$$I = const \cdot \left|F(Q)\right|^2 , \qquad (1)$$

where the structure factor of the superlattice for (00*l*) scans, $F(Q)$, is written as

$$F(Q) = F(00l) = \sum_{n=0}^{N-1} F_0 e^{iQ\Lambda Q} . \qquad (2)$$

Q is the scattering vector along the (001) direction, Λ is the superlattice wavelength and *N* is the number of periods. $F_0$ is the structure factor of a single period and is given as

$$F_0 = \sum_{m=0}^{M-1} f_A e^{iQmd_A} + e^{iQ(Md_A + \delta)} \sum_{m=0}^{M-1} f_B e^{iQmd_B} e^{iQMd_A} + f_{IF} e^{iQ(Md_A + Md_B + \delta)} e^{iQd_{IF}} , \qquad (3)$$

where δ accounts for the change of layer spacing due to change of chemical bonds at the interfaces. *M* is the total number of monolayers in each superlattice layer and $f_A$, $f_B$, and $f_{IF}$ are the structure factors of the individual "InAs", "GaSb", and interfacial "GaAs" layers, respectively. The best fits to the data, shown as the solid lines in Figures 4, were obtained by using $InAs_{1-x}Sb_x$ and $GaAs_ySb_{1-y}$ alloyed layers rather than pure InAs and GaSb. This alloying is due to the existence of cross-contamination and segregation of As and Sb respectively, which has been observed in previous studies of InAs/GaSb superlattices. The fitting took into account that the MEE growth technique produces 95% of the desired interfacial bonds[10,13].

From the fit to sample D in Fig. 4a, the Sb percentage, $x$, in the InAs layer and the As percentage, $y$, in the GaSb layer, are 0.20 and 0.10, respectively. Previous works report that the cross-contamination in InAs/GaSb can be more than 30%[14,15] for both layers. From the fit to sample B grown on an InAs substrate, Fig. 4b, the $x$ and $y$ values for the superlattice layers are 0.12 and 0.05, respectively. The data for sample B is similar to that of sample D, except the substrate and the satellite peak widths increase with superlattice satellite order for sample B. This is a characteristic feature for superlattices with slightly fluctuating layer thickness. This peak broadening can be treated using the Hendricks-Teller approach[16,17]. Here, for simplicity, the peak broadening is fit by convolving the profile with a Gaussian function with a variable width as a function of $Q_z$.

B. Nanowire Samples

To quantitatively analyze the nanowire samples, higher ordered satellites and a low signal/noise ratio is needed in the XRD data; therefore, sample A was analyzed at NSLS. The intensity distribution around the GaSb $(\bar{2}24)$ reciprocal lattice point is shown in Fig. 6a. Both higher order vertical and lateral satellite peaks were observed. This data was simulated using a structural model based on the super *cr* unit cell marked by bright dots in Fig. 2a. The diffracted x-ray intensity is given by

$$I(Q_x, Q_z) = const \cdot \left| F(Q_x, Q_z) \sum_m \sum_n \delta(Q_x - mG_x)\delta\delta({}_z - nG_z) \right|^2 , \qquad (4)$$

where $\mathbf{Q}=(Q_x,Q_z)$ is the momentum transfer. $G_x$ and $G_z$ are defined as $G_x = 2\pi/\Lambda_x$ and $G_z = 2\pi/\Lambda_z$, with $\Lambda_x$ and $\Lambda_z$ being the periodicities of the nanowire array along the lateral (x) and vertical (z) directions, respectively. $F_{u.c.}(\mathbf{Q})$ is the structure factor of the super *cr* unit cell which is expressed as

$$F_{u.c.}(Q) = f_w(Q)[1 + e^{i(Q_x \Lambda_x/2 + Q_z \Lambda_z/2)}] \,. \quad (5)$$

The scattering amplitude, $f_w(\mathbf{Q})$, of a single nanowire was calculated based on the shape of the InAs wire (dark area) surrounded by a GaSb spacer (bright areas) as seen in Fig. 2a, using the equation

$$f_w(Q) = \int \sigma(r)\rho_w(r)e^{iQ\cdot(r+u)}dr \,. \quad (6)$$

Here, the shape function $\sigma(\mathbf{r})$ is equal to 1 if $\mathbf{r}$ falls inside the wire and 0 elsewhere. $\rho_w(\mathbf{r})$ is the electron density function of a single nanowire. The displacement field, $\mathbf{u}(\mathbf{r})$, which depends on $\sigma(\mathbf{r})$ and the composition, is related to the misfit strain fields $\varepsilon_{xx}$, $\varepsilon_{zz}$ and $\varepsilon_{xz}$ by

$$\varepsilon_{xx} = \frac{\partial u_x}{\partial x} = \frac{a_x - a_{sub}}{a_{sub}}, \quad \varepsilon_{zz} = \frac{\partial u_z}{\partial z} = \frac{a_z - a_{sub}}{a_{sub}}, \quad \varepsilon_{xz} = \frac{1}{2}\left(\frac{\partial u_x}{\partial z} + \frac{\partial u_z}{\partial x}\right), \quad (7)$$

where $a_x$ and $a_z$ are the in- and out-of-plane lattice constants of the strained film, and $a_{sub}$ is the lattice constant of the substrate.

The elastic strain and displacement field were determined by the solution of the elastic equilibrium equation without the presence of the volume forces, $\frac{\partial \sigma_{jk}}{\partial x_k} = 0$, where $\sigma_{jk}$ is the stress tensor. In linear elasticity, the stress tensor components correspond to the strain tensor components by Hooke's law, $\sigma_{jk} = C_{jklm}\varepsilon_{lm}$, where $C_{jklm}$ is the elastic constant tensor and the strain tensor, $\varepsilon_{lm}$, is evaluated with respect to the lattice of the unstrained material. If the strain tensor with respect to the substrate is used, one obtains

$$\sigma_{jk} = C_{jklm}\left(\varepsilon_{lm} - \delta_{lm}\varepsilon_{(M)}\right), \qquad (8)$$

where $\varepsilon_{(M)} = \dfrac{a_{(M)} - a_{sub}}{a_{sub}}$ is the lattice mismatch of the M-th layer with respect to the substrate. The complete system is given by the boundary conditions on the surface and the interfaces. The boundary conditions represent the continuity of the displacement on the interfaces, $u_j^{(-)} = u_j^{(+)}$, where the indexes (+) and (-) denote the values obtained from the upper and lower layer, respectively. The other condition is that the tractions acting on the boundary are in equilibrium, $t_j^{(-)} = -t_j^{(+)}$[18]. The surface traction is expressed from the stress tensor as $t_j = \sigma_{jk} n_k$, where ***n*** is the outward surface normal. The traction can also be expressed as a function of the strain tensor using equation (8) as $t_j = C_{jklm}\varepsilon_{lm} n_k - C_{jklm}\delta_{lm}\varepsilon_{(M)} n_k$, where we denote the second term as $t_j^{\infty}$. In the isotropic continuum, $t_j^{\infty}$ has a simpler form given as $t_j^{\infty} = E/(1-2\nu)\varepsilon_{(M)} n_j$, where E is the Young modulus and ν is the Poisson ratio. On the stress free surface the surface traction, ***t***, equals zero $t_j\big|_{surface} = 0$. Since the multilayer is much thinner than the substrate, we can assume that the substrate, far below the multilayer, is undeformed with $u\big|_{z\to-\infty} = t\big|_{z\to-\infty} = 0$. We have used the boundary integral (BI) method based on the Somigliana's integral equation, which is the integral formulation of the elastic equilibrium equation[19]

$$c_{jk} u_k(x_0) = \int_S \left\{U_{jk}(x - x_0)\left[t_k(x) + t_k^{\infty}(x)\right] - T_{jk}(x - x_0) u_k(x)\right\} dx, \qquad (9)$$

where $c_{jk}(x_0)$ equals $1/2\delta_{jk}$ for $x_0 \in S$, $u_j(\boldsymbol{x})$, $t_j(\boldsymbol{x})$ are the *j*-th components of the displacement and surface traction in position ***x***, respectively, and $U_{jk}(\boldsymbol{x}-\boldsymbol{x}_0)$ and $T_{jk}(\boldsymbol{x}-\boldsymbol{x}_0)$

are the isotropic Green's functions of the two-dimensional periodic structure which were presented by Yang and Srolovitz[19]. The integration region must contain material with constant lattice mismatch; therefore, each layer was a separate integration region. Since the neighbor layers are related by the boundary conditions, a coupled set of the integral equations must be solved numerically.

The interfaces and superlattice layer surfaces were divided into equidistant elements assuming that the displacement and traction were constant inside each element. Somigliana's equation was transformed into a set of the linear algebraic equations using the boundary conditions[18]. The particular integrals of the Green's functions over the single elements were calculated by the Gaussian quadrature with 20 points per element. If $\boldsymbol{x}$ is equal to $\boldsymbol{x}_0$, the functions $U_{jk}(\boldsymbol{x}-\boldsymbol{x}_0)$ and $T_{jk}(\boldsymbol{x}-\boldsymbol{x}_0)$ become singular. In such cases we have used the technique which allows us to calculate the singular part of the integrals analytically and the nonsingular part by the Gaussian quadrature[20].

Similar methods have proven to be valid for layers as thin as one monolayer[21]. For example, Fig 7 shows the calculated distribution of normal strain $\varepsilon_{xx}$ and $\varepsilon_{zz}$ in a single super *cr* unit cell of the nanowires. $\varepsilon_{xx}$ is almost uniform in the $InAs_{0.8}Sb_{0.12}$ and $GaAs_{0.05}Sb_{0.95}$ layers as well as at the interfaces. Alloying in the layers due to contamination/segregation will be explained later. However, the magnitude of the misfit strain at the interface is an order of magnitude higher than that in the bulk film. Also, $\varepsilon_{xx}$ fluctuates in a small range of about $5\times10^{-4}$ which is actually beyond the resolution of our experiment, but overall $\varepsilon_{xx}$ in the $InAs_{0.88}Sb_{0.12}$ layer is positive, indicating that the layer is slightly relaxed.

For sample A, the simulation, Fig. 6b, using equation (4) yields the essential features of the experimentally measured map shown in Fig. 6a; however, this is not enough to make a quantitative comparison. To do so, the intensity profile along $Q_x$ at $Q_z$ = 3.9887 r.l.u. has been extracted from the RSM and is shown in Fig. 8a. The satellites are indexed with two integers, (*m*,*n*), corresponding respectively to their lateral and vertical orders, respectively. Because of the *cr* symmetry, only the even-order peaks can be seen in this profile. The best fit, the solid line in Fig. 8a, was obtained by taking an average in-plane lattice constant of 6.1005 Å for both the "InAs" (whose actual composition is discussed below) and the InSb layers.

To determine the out-of-plane lattice constant, a $Q_z$-scan around the GaSb (004) reciprocal lattice point, Fig. 9, was performed. Here the GaSb substrate peak is resolved and is used as a reference for the determination of the out-of-plane lattice constant of the superlattice layers. The zeroth order satellite peak, located at $Q_z$ = 3.9887 r.l.u., yields an average out-of-plane lattice constant of 6.1134 Å for the superlattice. This value is larger than the expected value, 6.0882 Å, which would be determined from the bulk values of the lattice and elastic constants for pure InAs, GaSb and InSb and assuming 100% InSb interfacial bonds.

With this lattice constant, either InAs, GaSb, or both must have an increased out-of-plane lattice constant. This is possible if cross-contamination of the group V atoms has occurred (similar to what was observed for the planar samples). Sb incorporation into InAs leads to an $InAs_{1-x}Sb_x$ alloy whose lattice constant increases with the Sb concentration; similarly, As incorporation in the GaSb leads to a $GaAs_ySb_{1-y}$ alloy whose lattice constant decreases with increase in the As concentration. Simulations of the ($\bar{2}24$

) and (004) $Q_z$ scans, Figs. 8(b) and 9, respectively, using $InAs_{0.88}Sb_{0.12}$ and $GaAs_{0.05}Sb_{0.95}$ alloys, which have average out-of-plane lattice constants of 6.1180 Å and 6.0539 Å, respectively, yield excellent fits to the data. A similar RSM about both azimuths, Fig. 10, was observed for nanowire sample C about the ($\bar{2}24$) reciprocal lattice point. The average vertical and lateral periodicities of the nanowires in this sample are 148 Å and 1693 Å, respectively. A $Q_x$-scan across the central peak, (0,0), shown in Fig. 11, shows an asymmetrical intensity distribution, which suggests that the alloyed InAsSb layers, with a larger lattice constant than that of GaSb, are partially strain relaxed. This requires that the Sb fraction in these layers is at least 9% as can be seen using Vegard's law to calculate the dependence of the average lattice constant of $InAs_{1-x}Sb_x$ on Sb fraction, Fig. 12.

V. Discussion

The "InAs" layers in the nanowire samples appear to follow the SK growth mode. The growth experiments indicate that a transition from 2D to 3D began at a thickness of 2-3ML's. This transition occurs because 3D growth will result in a relief of strain energy greater than the increase in surface energy and is thus thermodynamically favored. InAs/GaAs (001) is a typical SK system with a misfit of 7% and a typical critical layer thickness of 1-2 ML's. Since strain energy is proportional to the square of the misfit strain, we would expect that the critical layer thickness for InAs to be about 100-200 ML's when it is grown on GaSb (001) because the misfit is reduced to about 0.62%. However, as mentioned earlier, the 2D to 3D growth transition of InAs in this system with InSb interfacial bonds was observed to start at about just 2-3 ML's. In the following, we will discuss, based on our x-ray results, why and how the critical layer thickness depended on the types of the interfacial bonds.

We first summarize our simulation results to find out the configurations of lattice misfit strain in our samples. We obtain the average freestanding lattice constant, $a_0$, of the superlattice layers by Vegard's law using the alloy composition determined from our simulation. The average misfit, *f*, is calculated as $f = (a_0 - a_{sub})/a_{sub}$. We calculated the average misfit of the two component layers and the interfacial bonds, which form a single atomic layer, of our samples. The results are listed in Table II, together with the misfit of pure InAs and GaSb with respect to the substrates used.

From this table, we see that the misfit of InAsSb layers of the two flat-layer superlattice samples has a sign opposite that of the GaAs interfacial monolayer. Conversely, for the two nanowire samples, A and C, the misfit of InAsSb layers has the same sign as the misfit of the InSb interfacial monolayer. Also noticed is that the misfit of the interfacial monolayer, InSb or GaAs, has a misfit an order of magnitude higher than that of the InAsSb layers. This suggests that the combination of the misfit strain of the interfacial monolayer and that of the InAsSb over-layer, including both their sign and magnitude, is crucial in determining the growth morphology of the InAsSb layers which eventually results in the self-assembly of the nanowires in Fig. 2(a).

Fig. 13 compares the strain configuration of InAsSb layers grown on top of InSb and GaAs interfacial bonds, respectively. Without Sb, the InAs layers would experience tensile strain because InAs (6.0584 Å) has a smaller lattice constant than GaSb (6.0961 Å). The InSb interfacial monolayers, however, experience a compressive strain because of their much larger lattice constant (6.4789 Å). Thus, the InAs layer and the InSb interfacial monolayer form a balanced strain configuration stabilizing the entire structure because their relaxation would require atomic displacements along opposite directions. In contrast, after a sufficient amount of Sb has been alloyed into InAs, both the InAsSb layer and the InSb interfacial monolayer will experience compressive strain forming an enhanced strain configuration, which makes them unstable; i.e., strain relaxation is more likely to occur.

The importance of the interfacial bonds can be seen more clearly by comparing samples D and A. Both of them were grown on GaSb substrates but with different interfacial bonds. Notice that the Sb fraction in the InAsSb alloy of the planar sample, D, is higher than that in the nanowire sample, A. However, the former one remains flat over its entire thickness because it is grown on top of GaAs bonds; whereas, the latter one turns into 3D growth because it is grown on top of InSb bonds. A similar conclusion can be drawn by comparing sample A with B which was grown on an InAs substrate with GaAs interfacial bonds. The use of InAs instead of GaSb as substrate increases the misfit strain of the InAsSb layers. Both samples have the same 12% Sb fraction in the InAsSb alloy layers thus the growth mode of these samples is not the composition effect. The InAsSb layers of sample A have a misfit of +0.21%, but the same InAsSb layers of sample B have a misfit of +0.83%. However, the observation is that the sample with a smaller misfit becomes unstable while the one with a larger misfit remains stable. The primary difference between the $InAs_{0.88}Sb_{0.12}$ layers in the two samples is that they were grown with different IF bonds. This clearly demonstrates that both the magnitude and sign of the misfit of the IF layer is key to the morphological instability of the $InAs_{0.88}Sb_{0.12}$ layers.

At this point, it is worthwhile to note that although our x-ray experiments were carried out with multilayer structures for the sake of enhanced x-ray signals, which are necessary for a quantitative analysis, our results are true for single-layer films as well. For example, Okada et al.[9] observed that the growth of $In_xGa_{1-x}As$ single-layer films on InP (001) substrates is stable if the films have a positive misfit of 0.5% ( $x = 0.605$ ) with respect to the substrate, but unstable if the films experience a –0.5% misfit ( $x = 0.45$ ). This is probably because the IF bonds in this system are dominantly GaP[22], which upon formation over the InP substrate, experience a misfit strain up to –7.12%, the stable and unstable growth of the films with positive and negative misfits is thus expected.

The reduction of the critical layer thickness for 2D to 3D growth transition due to interfacial bonds can be understood by employing a simple thermodynamic model. We will follow the approach of Sasaki et al.[23] who used it to explain the transition thickness of some SK systems with large misfit; such as the common atom system of InAs/GaAs (001) and Ge/Si (001). For our non-common atom system, however, we need to add the interfacial bonds into consideration.

Consider the morphologies of a film before and after instability transition as shown in Figs. 14(a) and (b), respectively. Formation of a triangular ridge surface (Fig. 14(b)), which represents a simplified version of the observed morphology from sample A, is due to elastic relaxation which lowers the strain energy but increases the surface energy. The strain energy stored in the film before relaxation is given by

$$U_{str} = \frac{E_i}{1-\nu_i}\varepsilon_i^2 lwt_i + \frac{E_f}{1-\nu_f}\varepsilon_f^2 lwt_f \,, \qquad (10)$$

where $l$, $w$, $t_i$, and $t_f$ are the length, width, thickness of the interfacial layer, and the thickness of the film, respectively. $E_x$, $\varepsilon_x$, and $v_x$ ($x = i, f$) are Young's moduli, strain and Poisson's ratios of the interface ($x = i$) and the film ($x = f$), respectively. The surface energy of the film is $U_{surf} = lw\lambda$, where λ is the surface energy per unit area of the planar surface. The strain energy after relaxation, $U'_{str}$, is expressed as

$$U'_{str} = \frac{1}{2}\left( \frac{E_i}{1-v_i} R_i^2 \varepsilon_i^2 lwt_i + \frac{E_f}{1-v_f} R_f^2 \varepsilon_f^2 lwt_f \right), \qquad (11)$$

where the terms $R_i\varepsilon_i$ and $R_f\varepsilon_f$ ($0 \le R_x \le 1, x = i or f$) are introduced to take into account the effective residual strain of the interface and film. This assumes that the volume of the film is conserved. The factor ½ appears because we consider that the relaxation is uniaxial. Note that although the strain in the ridged film will have a distribution, only the total residual strain energy is relevant to the problem; thus, we can always find the effective residual strain, $R_f\varepsilon_f$, for the film after relaxation.

The surface energy of the pyramid, $U'_{surf}$, is $U'_{surf} = lw\lambda'/\cos\theta$, where $\theta$ is the angle between the base and the side face of the pyramid and λ′ is the surface energy density of the pyramidal surface.

The total energy of the system before and after relaxation is then $U_{bfr} = U_{str} + U_{surf}$ and $U_{aft} = U'_{str} + U'_{surf}$, respectively. If $U_{bfr} < U_{aft}$, the film is stable against perturbation, whereas if $U_{bfr} > U_{aft}$, the film is unstable. The transition thickness is thus determined by $U_{bfr} = U_{aft}$ to be

$$t_{c,f} = \frac{2(\lambda' - \lambda\cos\theta)(1-\nu_f)}{E_f \cos\theta\, \varepsilon_f^2 (2 - R_f^2)} - \frac{\varepsilon_i^2 (2 - R_i^2) E_i (1-\nu_f)}{\varepsilon_f^2 (2 - R_f^2) E_f (1-\nu_i)} t_i \ . \tag{12}$$

The first term in Eq. (12) contains parameters of the film only, and is the intrinsic transition thickness of the film when interfacial bonding is the same as that within the film, and is thus equivalent to the transition thickness of the common-anion system[23]. The second term results in the decrease in the transition thickness of the film due to the presence of the interfacial layer. If $\varepsilon_i$ is of the same order of magnitude as $\varepsilon_f$, such a reduction will be minimal since the interface thickness, $t_i$, is a single monolayer (about 3 Å for most III-V semiconductors). However, if $\varepsilon_i$ is an order of magnitude higher than $\varepsilon_f$, the transition thickness can be reduced by a few hundred monolayers, or hundreds of angstroms, which is quite significant. Moreover, a smaller $R_i$, in comparison with $R_f$, is expected as a guarantee that the second term is always smaller than the first one because $t_{c,f}$ cannot be negative. This is true because the relaxation of a pyramid is always the lowest at the base and the highest at the top. We should also note that Eq. (12) is valid only if $\varepsilon_f$ and $\varepsilon_i$ are of the same sign. If $\varepsilon_f$ and $\varepsilon_i$ are of opposite sign, the large interfacial strain will not lead to Eq. (12) because the relaxation of the interfacial layer and the layers on top of it will involve atomic displacements along opposite directions; i.e., they are in a strain-balancing configuration, which actually prohibits the relaxation from occurring[24].

## VI. Summary

We have performed detailed x-ray analyses of a series of InAs/GaSb superlattices grown by MBE with either InSb or GaAs interfacial bonds. The morphologies of the films show strong dependence on the types of the interfaces. We found that if the large misfit at the interface is of the same sign as that of the small misfit of the over-grown layer, morphological instability can be triggered shortly after the beginning of growth of the over-layers. However, if the interfacial layer and the over-layer experience an opposite misfit, they are stabilized. Our findings indicate that with proper design of the interfacial bonding, self-assembled nanostructures can be grown in material systems with small misfit, which would otherwise be impossible. This explains the occurrence of the morphological instabilities observed in the epitaxially grown InAs/GaSb superlattice system, which has a variety of potential applications as lasers and detectors operating in the mid- to far-infrared region (3-14 μm)[25-27]. Therefore, our results demonstrate an approach that may be useful for creating a novel class of technologically important semiconductor nanostructures.

Acknowledgement

The authors are grateful to B.Z. Nosho, B.R. Bennett and L.J. Whitman for supplying the samples. The work is supported by the NSF through grants DMR-0237811 (DWS), DMR-0406323 and DMR-0908286 (KEB), and DMR-0408539 (SCM), and by the Texas Center for Superconductivity and Advanced Materials.

Figure Captions

Fig. 1. Schematic diagrams of atomic structure of an InAs/GaSb superlattice with (a) GaAs and (b) InSb interfacial bonds.

Fig. 2. Reconstructed 3D structures of (a) sample A and (b) sample B based on cross-sectional STM images taken on both the (110) and $(\bar{1}10)$ planes demonstrating the formation of a 2D nanowire array with centered-rectangle (*cr*) symmetry in the unstable sample. The dark areas are InAs and the bright areas are GaSb and the basic building block of this structure, i.e. a single nanowire, is delineated by solid lines in (a).

Fig. 3. Diffraction geometries used for the XRD measurements. (a) For azimuth 1, the plane of diffraction is parallel to [110] direction and (b) for azimuth 2, the plane of diffraction is parallel to $[\bar{1}10]$ direction.

Fig. 4. A scan around the GaSb (004) reciprocal lattice point along [001] direction for (a) planar sample D and (b) planar sample B, grown with GaAs interfaces. The dots and line are experimental and calculated curves, respectively. The peak broadening for sample B is treated by the Hendricks-Teller approach [15,16].

Fig. 5. The measured (002) reciprocal pace map of sample A (a) parallel to the [110] direction and (b) parallel to $[\bar{1}10]$ direction.

Fig. 6. The (a) measured and (b) simulated RSM sample A around GaSb $(\bar{2}24)$ reciprocal lattice point.

Fig. 7. The calculated strain fields, $\varepsilon_{xx}$ and $\varepsilon_{zz}$, for a single nanowire.

Fig. 8. The line profiles extracted from the map, as indicated by the lines in Fig. 6., for sample A along the $[\bar{1}10]$ and [001] directions, respectively. The dots and lines correspond to the measured and calculated curves, respectively.

Fig. 9. The (004) line scan along the [001] direction of sample A. The high resolution setups allow clear separation of the substrate and the central superlattice peaks.

Fig. 10. The (224) and $(\bar{2}24)$ reciprocal space maps of sample C taken with the two geometries in Fig. 3.

Fig. 11. The line profile along $[\bar{1}10]$ direction extracted from Fig. 10, as directed by the dotted line there. The asymmetry of the satellite intensities is similar to that of Fig. 8(a).

Fig. 12. The lattice parameter of $InAs_{1-x}Sb_x$ alloy as a function of Sb composition according to Vegard's law compared with that of GaSb. A cross point is seen at about x = 0.09 above which the alloy's lattice becomes larger than that of GaSb.

Fig. 13. The strain configurations of the InAs layer and the InSb interface. (a) No Sb is incorporated in InAs. (b) Sufficient Sb has been incorporated into InAs so that strain in this film changed from tensile to compressive.

Fig. 14. The morphologies of a III-V heterostructure before and after the SK instability transition.

Table I. Sample parameters for InAs/GaSb superlattices.

| Sample | Substrate | $T_{Growth}$ ($^oC$) | Interface (IF) | # of periods | Stability |
|---|---|---|---|---|---|
| A | GaSb | 405 | InSb | 100 | unstable |
| B | InAs | 395 | GaAs | 100 | stable |
| C | GaSb | 395 | InSb | 90 | unstable |
| D | GaSb | 395 | GaAs | 90 | stable |

Table II. Composition determined from x-ray data and the lattice misfit, *f*, of the superlattice layers.

| | Alloyed layers | | | | | |
|---|---|---|---|---|---|---|
| Sample | $InAs_{1-x}Sb_x$ | | $GaAs_ySb_{1-y}$ | | Interface (IF) | Stability |
| | x | *f*(%) | x | *f*(%) | | |
| A | 0.12 | +0.21 | 0.05 | -0.36 | InSb | unstable |
| B | 0.20 | +0.76 | 0.10 | -0.73 | GaAs | stable |
| C | >0.09 | >0.00 | - | - | InSb | unstable |
| D | 0.12 | +0.83 | 0.05 | -0.26 | GaAs | stable |

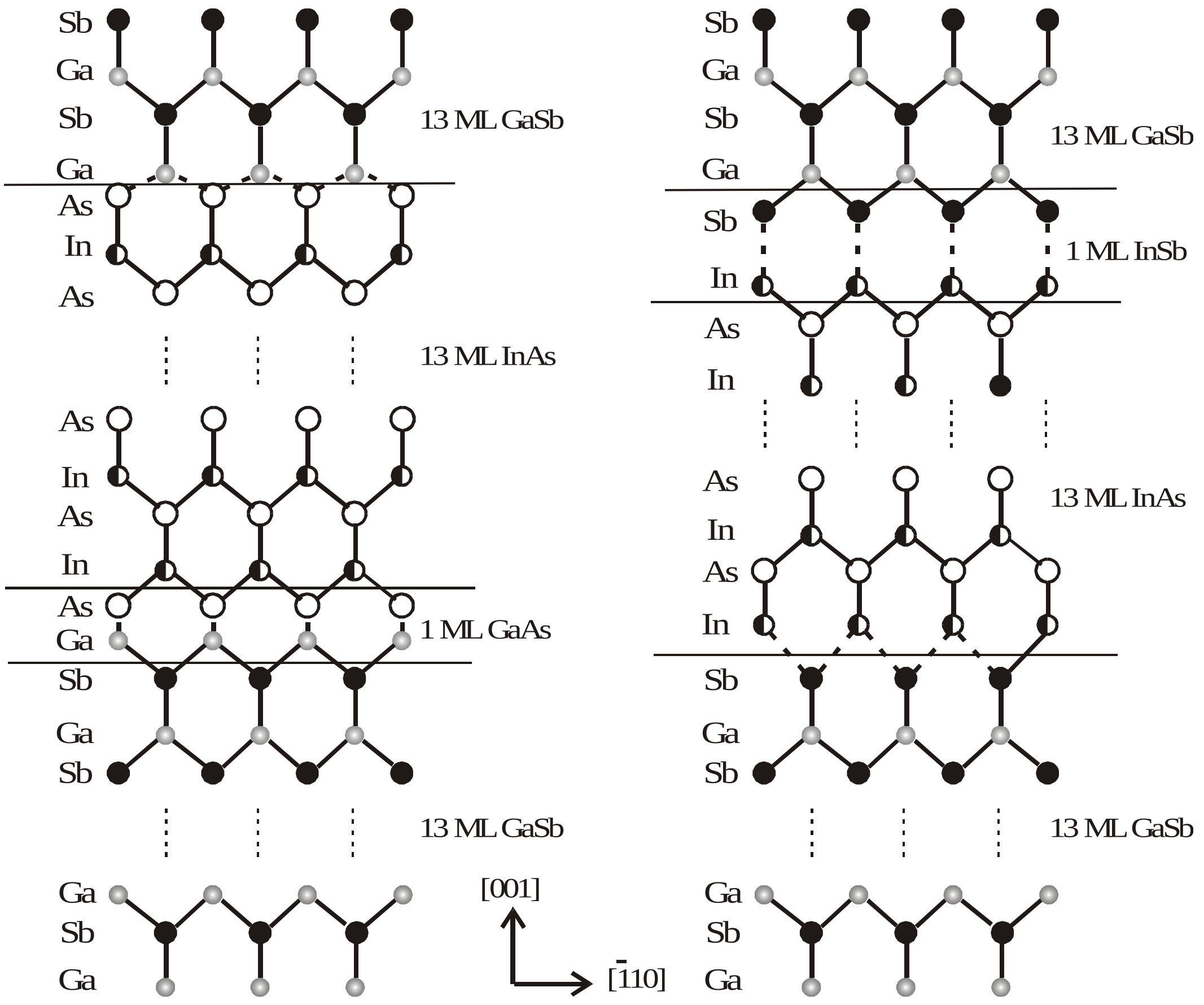

Fig. 1 Li et al.

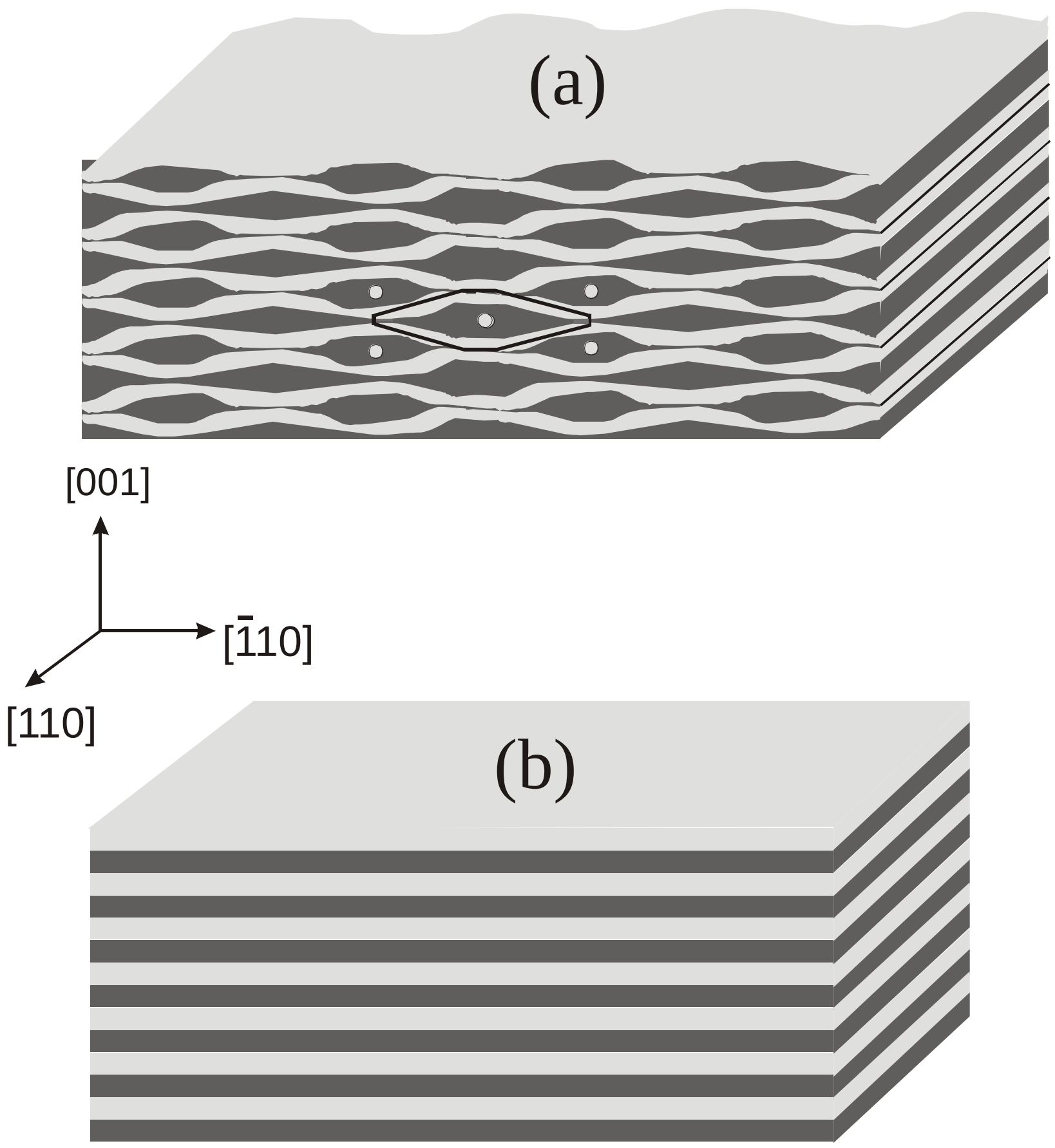

Fig. 2 J.H. Li et.al.

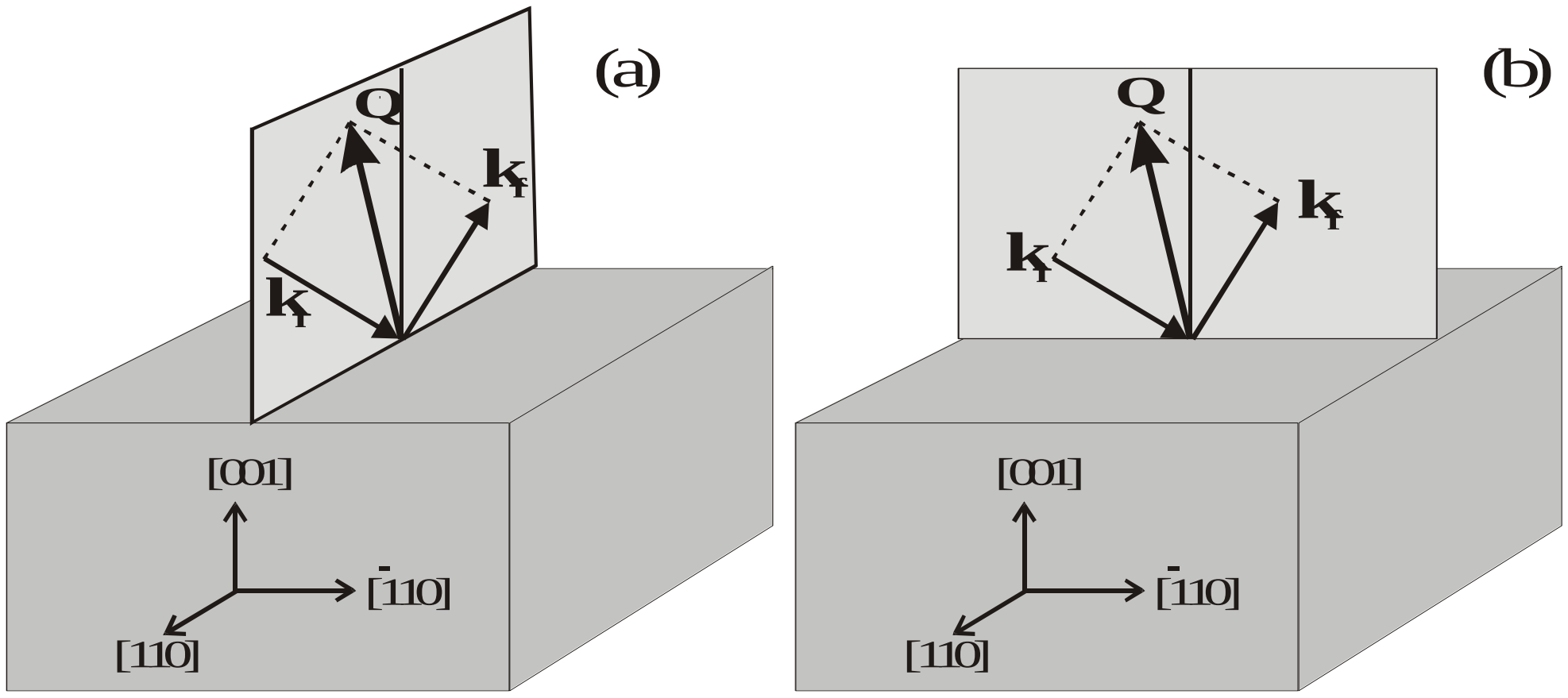

Fig. 3 Li et al.

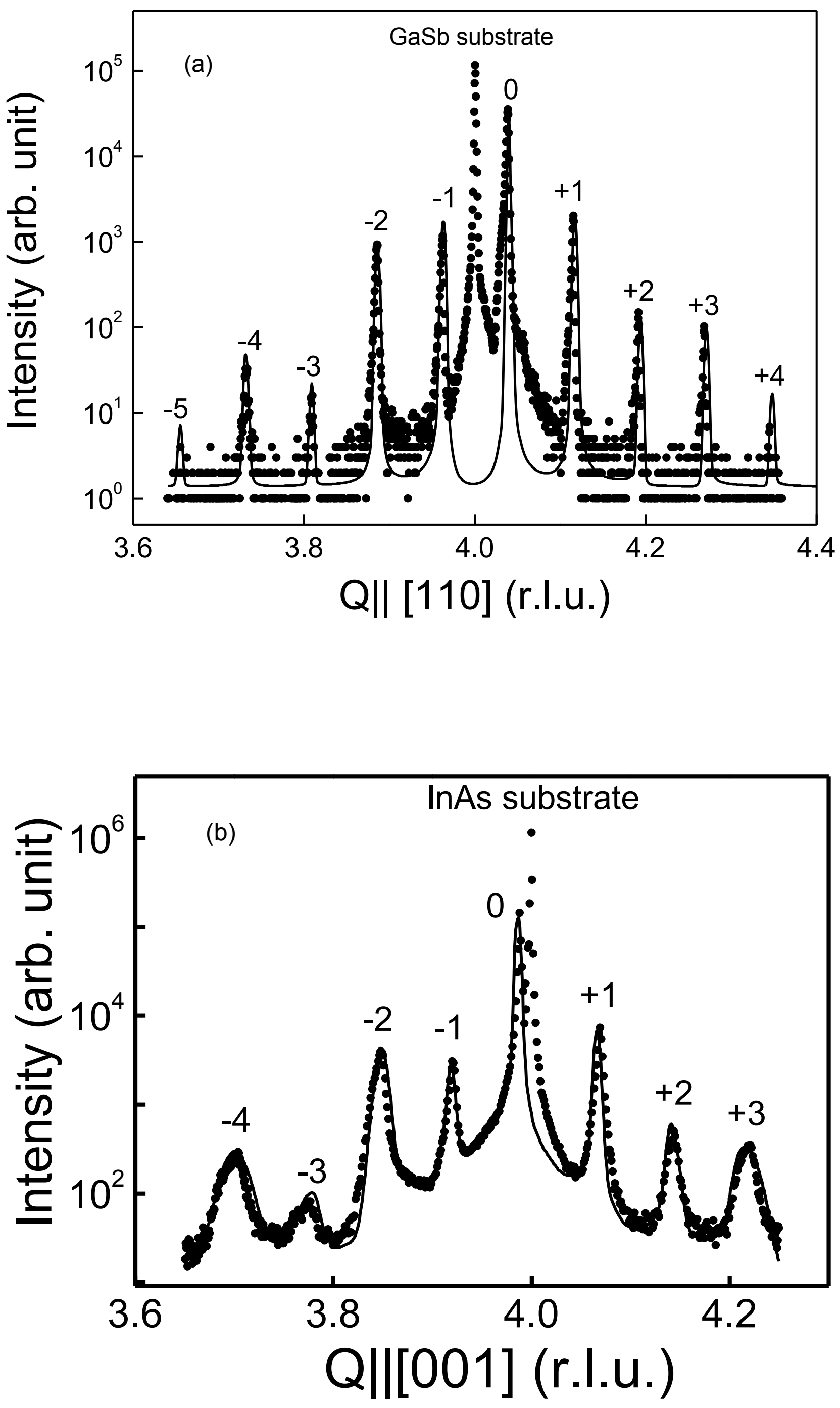

Fig 4 Li et al.

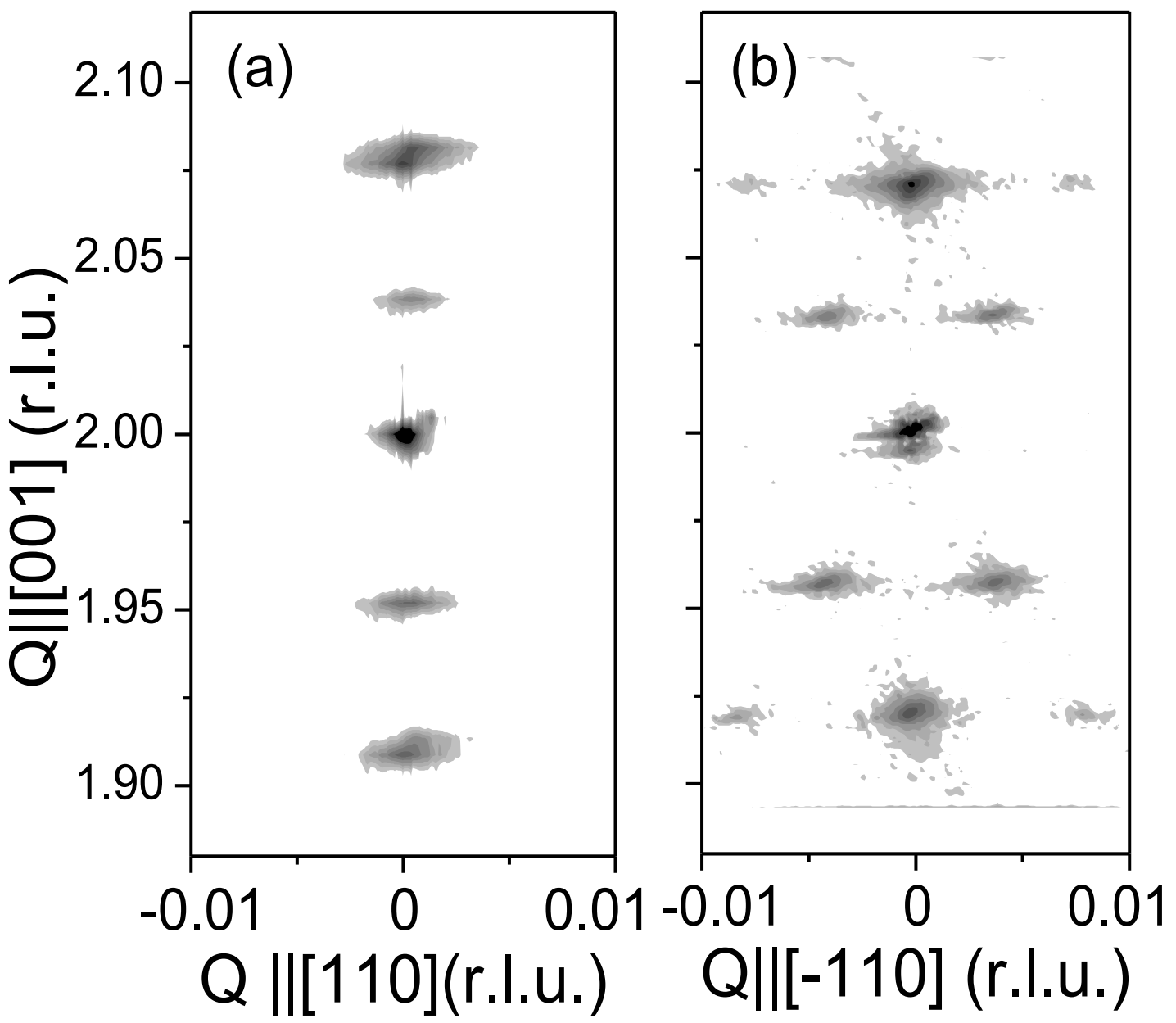

Fig. 5 Li et al.

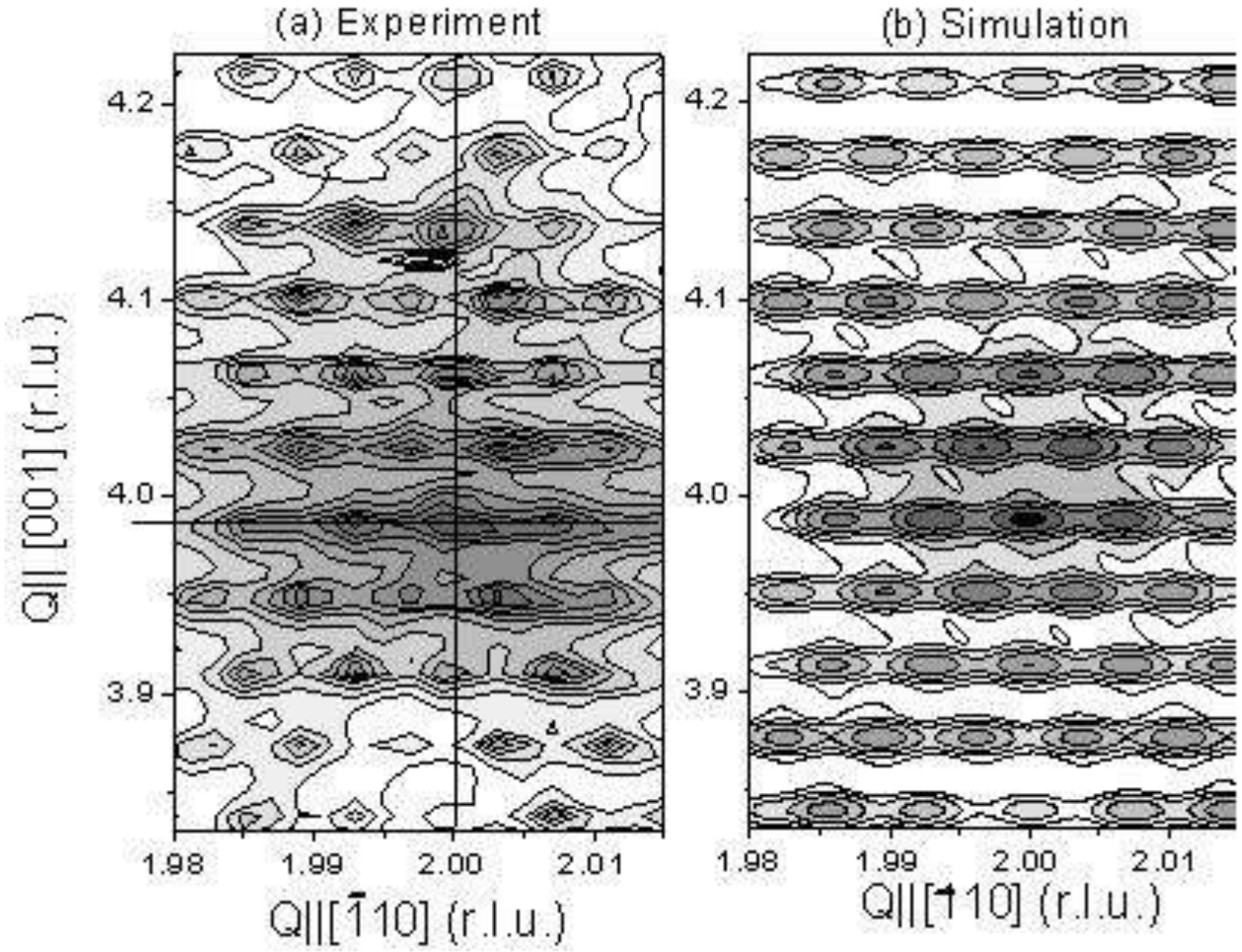

Fig. 6 J.H. Li et.al.

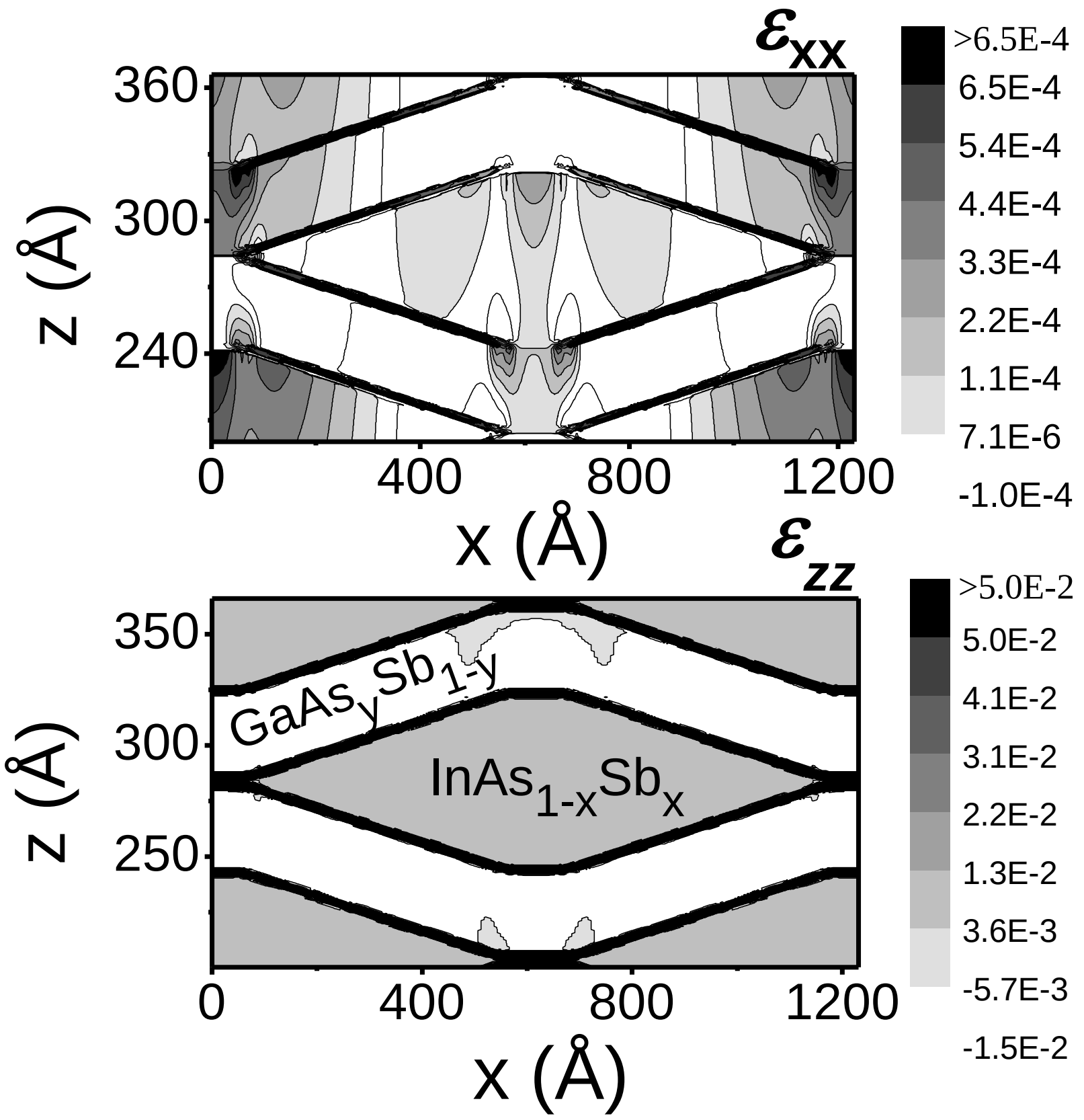

Fig. 7 Li et al.

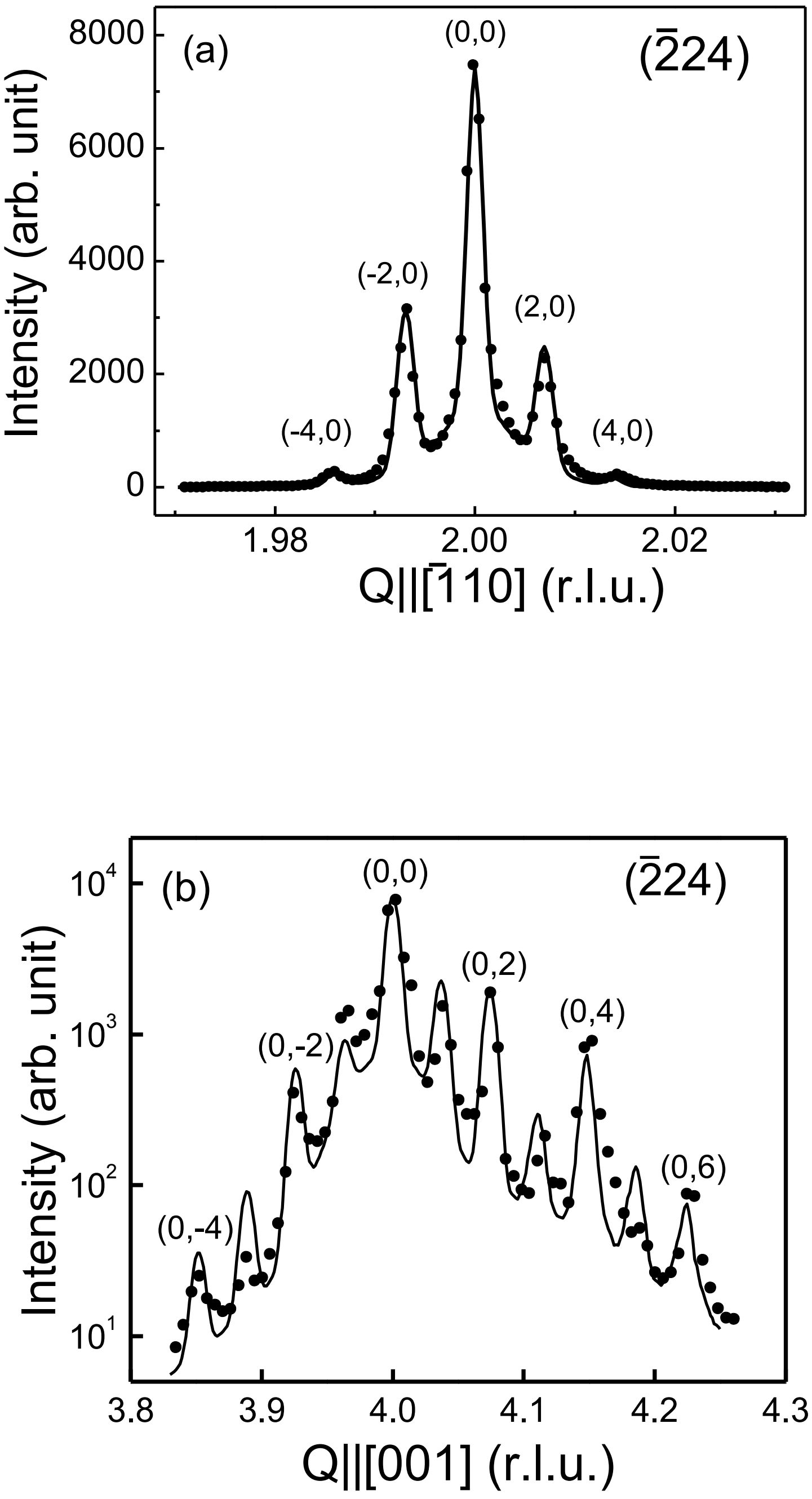

Fig. 8 Li et al.

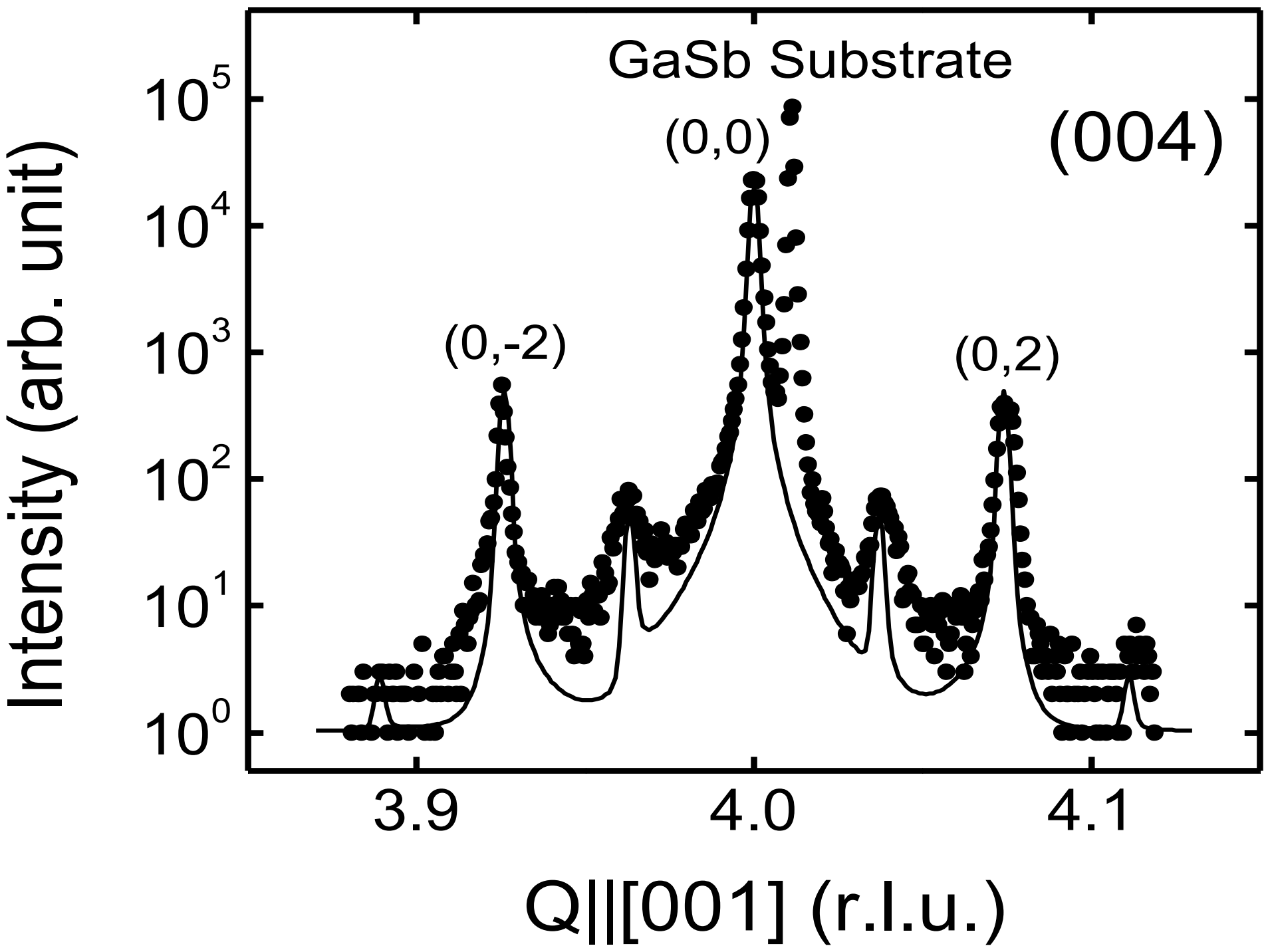

Fig. 9 J.H. Li et.al.

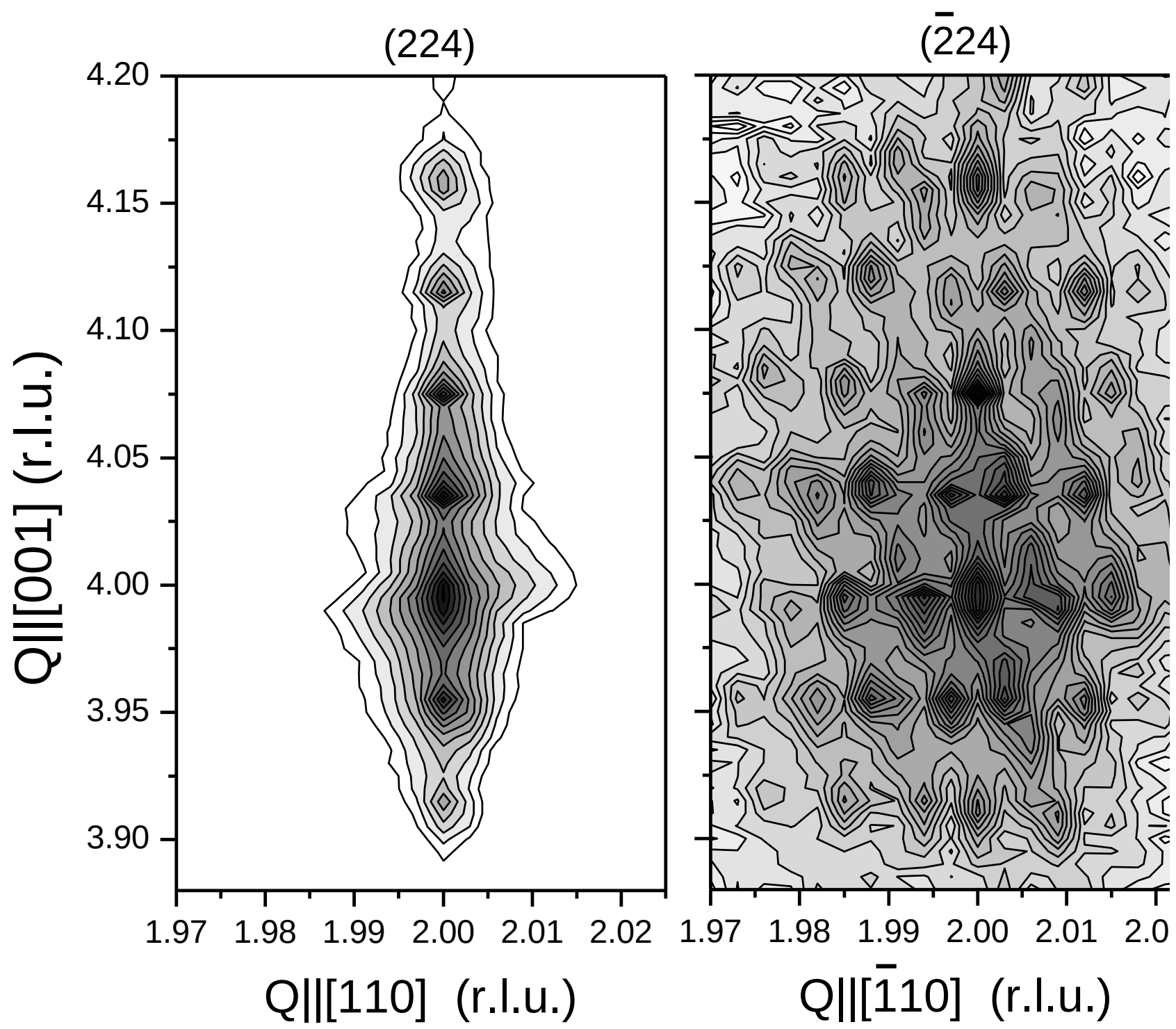

Fig. 10 J.H. Li et.al.

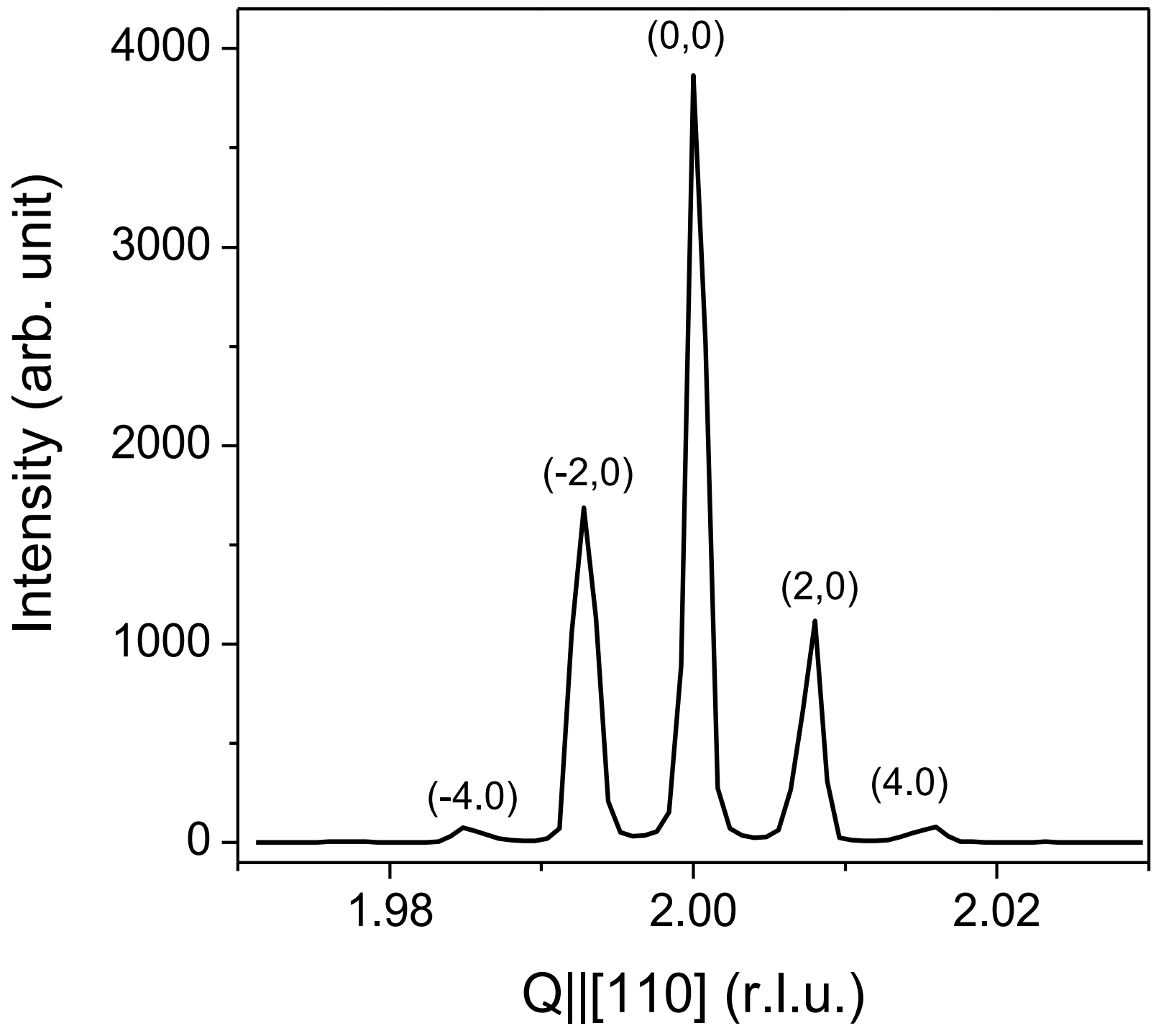

Fig 11 Li et al.

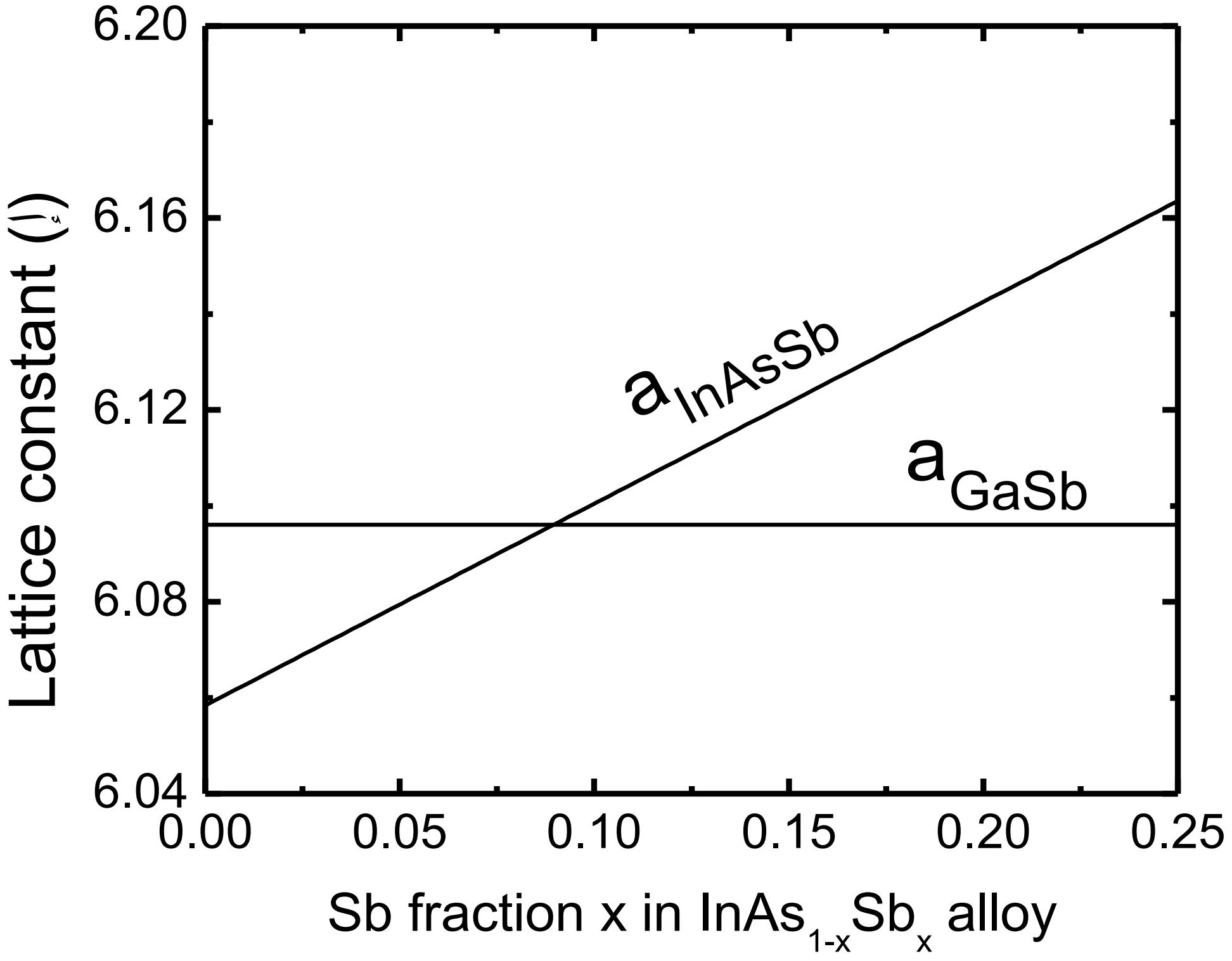

Fig 12 Li et al.

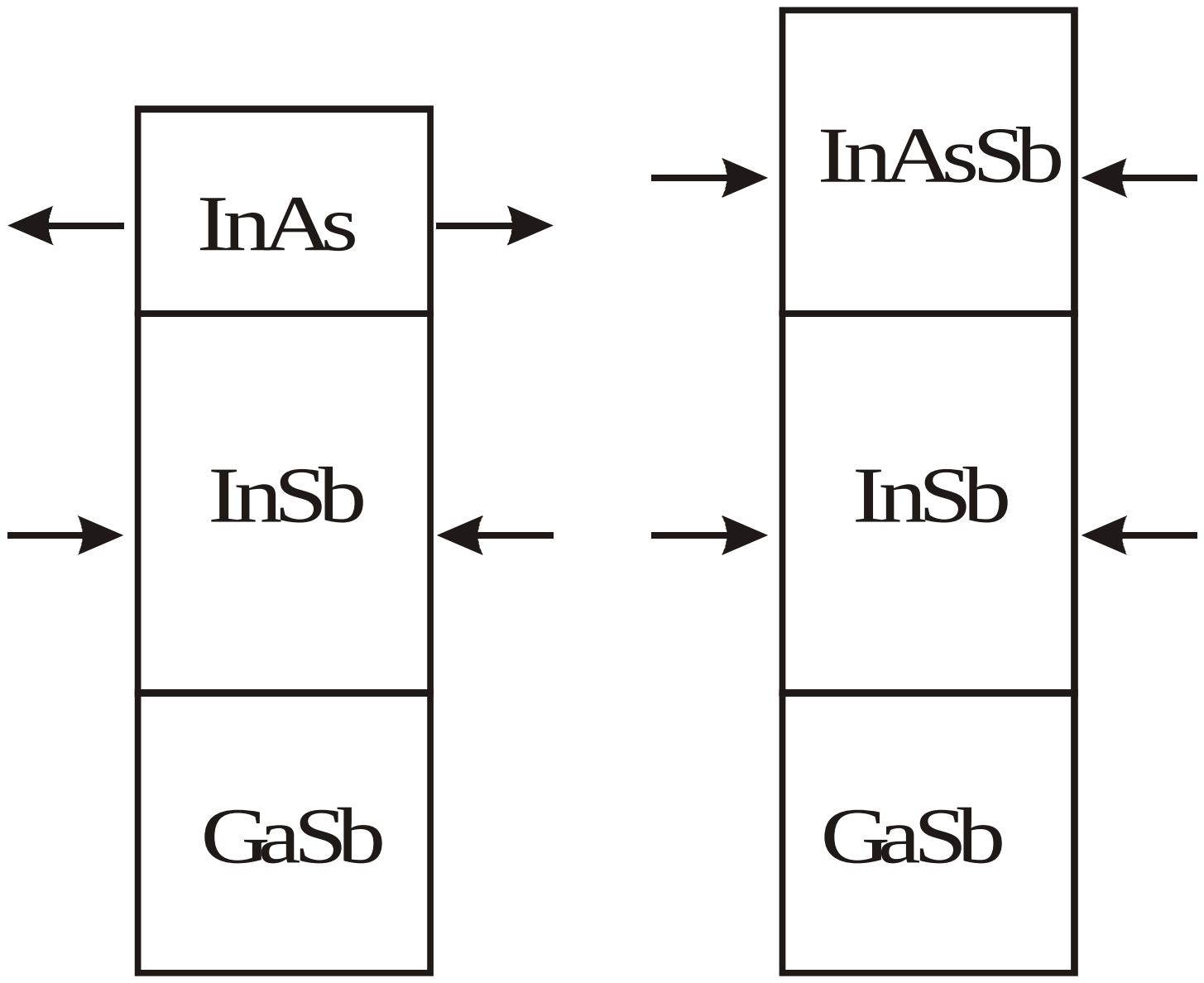

Fig 13 Li et al.

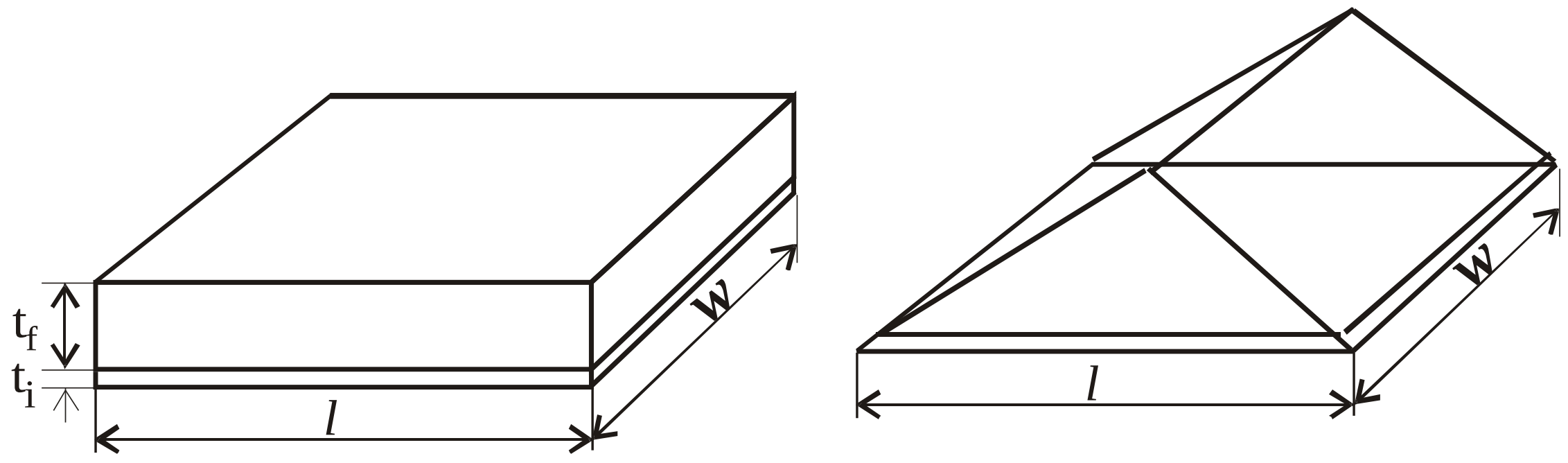

Fig 14 Li et al.